\documentclass[12pt]{article}
 
\usepackage{epsf}
\usepackage{graphicx}

\def\phfl{phot$\cdot$cm$^{-2}$s$^{-1}$ }
\def\*{$^{*}$}

\def\ergs{erg s$^{-1}$}

\def\Granat{\hbox{\it GRANAT}}
\def\deg{$^\circ$}

\def\arcsec{$^{\prime\prime\,}$}
\def\arcmin{$^{\prime\,}$}
\def\SLX{\mbox{SLX1744-299/300}}
\def\fdg{\hbox{$.\!\!^\circ$}}
\def\型{$\pm$}
\begin{document}

{\it To be published in the proceedings of ``X-Ray Astronomy 2000'',
Palermo, Italy}

\bigskip

\large
\centerline{\bf X-RAY BURSTERS A1742-294 AND SLX1744-299/300}
\centerline{\bf NEAR THE GALACTIC CENTER}
\vspace{15mm}

\normalsize
\centerline{A.A.Lutovinov$^{1,3,*}$, S.A.Grebenev$^{1}$, M.N.Pavlinsky$^{1}$,
R.A.Sunyaev$^{1,2}$}  
 
\vspace{5mm}
 
\noindent
$^1$ {Space Research Institute, Profsoyuznaya 84/32, 117810 Moscow, Russia}\\ 
$^2$ {Max-Planck-Institut f\"ur Astrophysik,
Karl-Schwarzschild-Str. 1, 85740 Garching bei \\ \noindent Munchen, Germany}\\
$^3$ {visiting INTEGRAL Science Data Center, ch. d'Ecogia 16, 1290 Vesoix,
Switzerland} \\
$^*$ {e-mail: aal@hea.iki.rssi.ru}

\vspace{7mm}

{\small We present a study of two X-ray bursters in the Galactic Center
field with the ART-P telescope onboard GRANAT. During 1990-1992
approximately 100 X-ray bursts were detected from the Galactic Center field
and most of them were identified with the two persistent sources, A1742-294
and SLX1744-299/300.  The burst recurrence times were found to be $\sim2.4$
and $\sim4.6$ hours, respectively, which is significantly shorter them was
previously thought. We calculated the mean and maximum fluxes during the
bursts and investigated their time profiles. We found that the shape of
A1742-294 bursts correlates with the burst peak flux. The spectral analysis
was carried out for the strongest burst, detected from A1742-294 on Oct 18,
1990.}

\vspace{10mm} 

\section{INTRODUCTION}

The first detailed X-ray map of the GC field was obtained by {\it EINSTEIN}
in the soft energy band ($\le4.5$ keV). In the harder band (4-30 keV) this
region was intensively studied only with the ART-P telescope, because other
missions ({\it Spacelab-2, Spartan-1} and telescope TTM on the {\it
MIR-KVANT} module) could not perform long observations for technical
reasons. The total ART-P/\Granat\ exposure time of the GC field observations
is $\sim$830 ks. Such a long exposure allowed to obtain the detailed X-ray
map of the GC region, to investigate in detail emission from persistent
sources, and to discover 4 new sources (Sunyaev et al., 1991, Pavlinsky et
al., 1992). 

This paper presents preliminary results of observations of two X-ray
bursters A1742-294 and \SLX\ located $\sim$1\deg from the GC. Both objects
are regular sources of type I bursts. The time profiles of this type bursts
depends on the energy band and are characterized by shorter decay time at
high energies. Type I bursts are thought to result from thermonuclear flares
on the surface of accreting neutron stars. \\

\section{OBSERVATIONS}
 
ART-P is a coded-mask telescope which is able to detect photons in the 3-60
keV energy band with the time resolution of 3.9 ms. The geometric area of
the position-sensitive detector is 625 cm$^{2}$; half of the detector is
covered with opaque mask elements. The telescope field of view is
3\fdg4$\times$3\fdg6, the energy resolution is $\sim$25\% at 5.9 keV and the
instrument dead time is about 580 $\mu$s. There are four identical
telescope units on \Granat. The telescope was described in detail by
Sunyaev et~al.(1990).  

The observations were carried out in the ``photon-by-photon'' mode, in which
coordinates, energy and arrival time of each photon were recorded into the
buffer memory. The data are transferred from buffer to the main satellite
memory when the buffer is filled. The data transfer happens each 150-200 sec
and takes $\sim$30 sec. This results in gaps between individual exposures in
ART-P data sets.\\

\section{RESULTS}

The telescope ART-P detected 100 x-ray bursts in the GC field during
1990-1992 (Grebenev et al., 2000). Lewin et al.(1976) found that most part
of X-ray bursts observed by {\it SAS-3} from the GC field originated from
three sources -- A1742-289, A1743-28 and A1742-294. During ART-P
observations no bursts were detected from A1742-289 and A1743-28. 

A1742-294 is the brightest persistent X-ray source in the GC field in the
standard X-ray energy band. It is responsible for $\sim1/3$ of the total
X-ray emission from this region. The source persistent flux in the 3-20 keV
energy band varied in the range 20--50 mGrab during ART-P observations. This
flux corresponds to a luminosity of $(0.4-1)\times10^{37}$ \ergs~ assuming a
distance of $\sim$8.5 kpc. In total, 26 type I X-ray bursts with a typical
duration of $\sim(15-20)$ sec were identified with A1742-294 (see Table 1,
fluxes are shown in the 3-20 kev energy band). In several observations, two
bursts from this source were detected. It allowed to measure the burst
recurrence time $t_r\sim$2.4~hr.  This value is several times smaller than
that obtained by Lewin et al.(1976).

The other GC field source for which ART-P detected X-ray bursts was \SLX.
This source was discovered by {\it Spacelab-2} (Skinner et al., 1987) and
{\it Spartan-1} (Kawai et el., 1988). In fact it is a double X-ray source,
with components SLX1744-299 and SLX1744-300 separated by $\sim$3\arcmin, of
which the latter is a burster. The angular resolution of ART-P,
$\sim$5\arcmin, does not allow to separate fluxes of these two sources. The
average flux from the double source measured by ART-P was equal to $\sim$20
mCrab, which corresponds to the luminosity $4.4\times10^{36}$ \ergs.  During
the ART-P observations 6 type I X-ray bursts were detected from this region
(see Table 2, fluxes are shown in the 3-20 kev energy band). One of these
bursts was extremely strong.  The coordinates of this burst
($R.A.(1950)=17^h44^m14.^s7$ and $Decl.(1950)=-29$\deg58\arcmin19\arcsec)
coincide with the SLX1744-299 position, while SLX1744-300 was outside the
position error cycle (Pavlinsky et al., 1994). The \SLX\ bursts recurrence
time was equal to $\sim$4.6 hr.

The energy spectra of the persistent emission of both sources, A1742-294 and
\SLX, are very similar. They can be described by a thermal bremsstrahlung
model with parameters in the range $kT_{\rm br}\simeq(6.4-10.5)$ keV for
A1742-294 and $kT_{\rm br}\simeq(4.7-9.2)$ keV for \SLX. The hydrogen column
density $N_{\rm H}$ was fixed at a value of $6\times10^{22}$ cm$^{-2}$.
Typical persistent spectra of these sources measured on Sept 9, 1990 are
shown in Fig.1. The energy spectra of A1742-294 bursts can be modeled by the
blackbody spectrum with temperatures in the range 1.4--2.7 keV; the mean
value of the temperature is $kT_{\rm bb}=1.81\pm0.39$ keV. The mean
black-body temperature for \SLX\ bursts was slightly higher, $kT_{\rm
bb}=2.40\pm0.32$ keV.

The time profiles of A1742-294 depend on the source flux during the burst.
The shape of the burst profile is almost ``triangle'', i.e. the rise time is
close to the decay time when the mean flux in the 3-20 keV energy was
$\sim$0.4\phfl. When the burst flux in the same energy band was higher,
$\sim$0.9\phfl, the burst profiles have a ``classical'' shape, i.e. a rapid
rise of intensity followed by a smooth decay. To study
this behavior in more detail, we constructed average profiles of weak and
strong bursts. We rejected bursts which occurred in the beginning or the end
of exposures. The bursts were normalized to the same peak intensity and
aligned by maxima of their intensity. In Fig.2, we show the average profiles
of weak and strong bursts in three energy bands. It is evident that the
shapes of weak and strong bursts in the hard energy band (8-20 keV) are
similar, while in the soft energy band (3-8 keV), the rise time of weak
burst is longer than that for strong bursts.

The average profile of 4 bursts from \SLX\ is presented in different energy
bands in Fig.3.

We carried out a spectral and timing analysis of the strongest X-ray burst
detected on Oct 18, 1990 from A1742-294. The burst peak flux was $\sim1.5$
Crab in the 3-20 keV energy band. The time profiles of this burst in four
energy bands are shown in Fig.4 (the zero point corresponds to the burst
maximum in the broad energy band 3-20 keV). It is clear that the burst
maximum at high energies is reached later than at low energies, i.e. a rise
time in the 12-16 keV energy band ($\sim$5 sec) is longer than that in the
4-8 keV energy band ($\sim$1-2 sec). On the contrary, the exponential decay
time decreases from 8.7 sec in the 4-8 keV energy band to 3.9 sec in the 8-12
keV band, 2.3 sec in the 12-16 keV band, and 1.5 sec in the 16-20 keV. No
significant burst emission was detected in the harder energy band ($\ge20$
keV). The evolution of the source X-ray luminosity in the 3-20 keV energy
band and its best-fit blackbody temperature are shown in Fig.5. It is
interesting to note that the highest temperature during the burst was
reached at the moment of maximum count rate at high energies.\\

\noindent     
This work was supported by RBRF grants 98-02-17056, 99-02-17178 and
00-15-99297.

\section{REFERENCES}

\noindent
Grebenev S., Lutovinov A., Pavlinsky M. et al. // 2000, in press.\\
\noindent
Kawai N. et al. // 1988, Astroph. J., {\bf 330}, 130.\\
\noindent
Lewin W., Hoffman J., Doty J. et al. // 1976, MNRAS, {\bf 177}, 83.\\
\noindent
Pavlinsky M., Grebenev S., Sunyaev R. // 1992, Sov. Astron. Lett., {\bf 18},
217.\\
\noindent
Pavlinsky M., Grebenev S., Sunyaev R. // 1994, Astroph. J., {\bf 425}, 110.\\
\noindent
Skinner G. et al. // 1987, Nature, {\bf 330}, 554.\\
\noindent
Sunyaev R. et al. // 1990, Adv. Space Res., {\bf 10}, 223.\\
Syunyaev, R., Pavlinskii, M., Churazov, E. et al. // 1991, Sov. Astr. Lett.,
{\bf 17}, 42. \\

\begin{table}[p]
\vbox{
\begin{minipage}{150 mm}
\centering
{\bf Table 1. }{\small X-ray bursts detected from A1742-294 by GRANAT/ART-P in
1990-1992.}\\   
\centering
\vspace{1mm}
\footnotesize{
\begin{tabular}{r|c|c|c|c|c|c}
\hline
 Date & $T_{0}$,&  Flux, & Flux max, & Duration, s  & Signifi-
& Head \\  
      &  UT     & \phfl  & \phfl    & & cance   &   \\
\hline
20.03.90  & $18^{h}05^{m}00^{s}$ & 0.379\型0.052&
                                   0.902\型0.092& 26 & 17.5 & 1 \\ 
          & 20 01 29& 0.311\型0.044& 0.962\型0.095& 18&16.4&1\\
24.03.90  & 18 32 42& 0.572\型0.068& 1.197\型0.109& 24&16.6&1\\
          & 21 42 00& 0.461\型0.059& 0.822\型0.086& 17&12.0&1\\
 8.04.90  & 14 06 38& 0.421\型0.053& 0.757\型0.080& 24&18.8&1\\
 9.09.90  & 14 07 48& 0.363\型0.041& 0.835\型0.071& 18&11.6&4\\
          & 15 53 21& 0.386\型0.043& 0.804\型0.069& 17&16.0&4\\
29.09.90  & 13 32 38& 0.164\型0.018& 0.279\型0.026& 30& 8.5&4\\
 5.10.90  & 15 31 44& 0.208\型0.030& 0.294\型0.039& 32&12.0&4\\
          & 19 43 33& 0.187\型0.029& 0.296\型0.040& 21&11.9&4\\
 6.10.90  & 22 10 36& 0.551\型0.057& 0.864\型0.074& 22&12.0&4\\
 9.10.90  & 17 20 52& 0.334\型0.038& 0.652\型0.060& 11&13.3&4\\
          & 19 06 11& 0.373\型0.042& 0.831\型0.070& 15&17.8&4\\
10.10.90  & 14 35 48& 0.369\型0.041& 0.608\型0.058& 12&15.3&4\\
          & 16 06 25& 0.589\型0.057& 0.798\型0.069&  9&12.5&4\\
18.10.90  &  9 50 31& 0.982\型0.078& 2.299\型0.125& 17&69.9&4\\
22.02.91  & 14 07 59& 0.987\型0.119& 2.055\型0.189& 13&33.6&3\\
23.02.91  & 22 24 36& 0.969\型0.119& 2.381\型0.207& 16&42.6&3\\
26.02.91  & 11 00 20& 0.911\型0.114& 2.026\型0.188& 15&19.5&3\\
 1.04.91  & 13 39 44& 0.507\型0.077& 1.090\型0.131& 28&16.4&3\\
          & 15 28 57& 0.705\型0.099& 1.413\型0.155& 19&18.5&3\\
 8.04.91  & 13 16 25& 0.654\型0.091& 1.601\型0.164& 22&13.0&3\\
15.10.91  & 19 59 08& 0.626\型0.095& 1.710\型0.178& 14&17.9&3\\
18.10.91  & 11 44 04& 0.592\型0.086& 1.321\型0.147& 15&24.6&3\\
21.02.92  & 12 52 53& 0.755\型0.103& 2.453\型0.213& 21&39.9&3\\
 2.03.92  &  1 43 22& 0.929\型0.120& 2.322\型0.207& 15&29.8&3\\

\hline
\end{tabular}}
\end{minipage}

\vspace{20 mm}

\begin{minipage}{165mm}

\centering
{\bf Table 2. }{\small X-ray bursts detected from SLX1744-299/300 by
GRANAT/ART-P in 1990-1992.}\\  
\centering
\vspace{1mm}
\footnotesize{

\begin{tabular}{r|c|c|c|c|c|c|c}
\hline
 Date & Source & $T_{0}$,&  Flux, & Flux max, & Duration, s  & Signifi-
& Head \\  
      &        &  UT     & \phfl  & \phfl      & & cance   &   \\
\hline

 9.09.90& \SLX   & 15 42 10& 0.339\型0.039& 0.627\型0.058& 19&12.2&4\\
        & \SLX   & 20 05 21& 0.529\型0.053& 0.940\型0.076& 16&13.8&4\\
10.09.90& \SLX   & 00 52 25& 0.270\型0.033& 0.657\型0.061& 15&17.2&4\\ 
 9.10.90& SLX1744-299& 19 19 58& 0.720\型0.064& 2.742\型0.138&170&86.3&4\\
18.10.90& \SLX   & 11 07 49& 0.583\型0.056& 1.211\型0.088& 21&17.1&4\\
23.02.91& \SLX   & 21 35 50& 0.391\型0.063& 0.930\型0.119& 15& 9.1&3\\ 

\hline
\end{tabular}}
\end{minipage}
}
\end{table}

\pagebreak

\begin{figure}[]
\epsfxsize=150mm
\hspace{20mm}\epsffile[100 400 590 700]{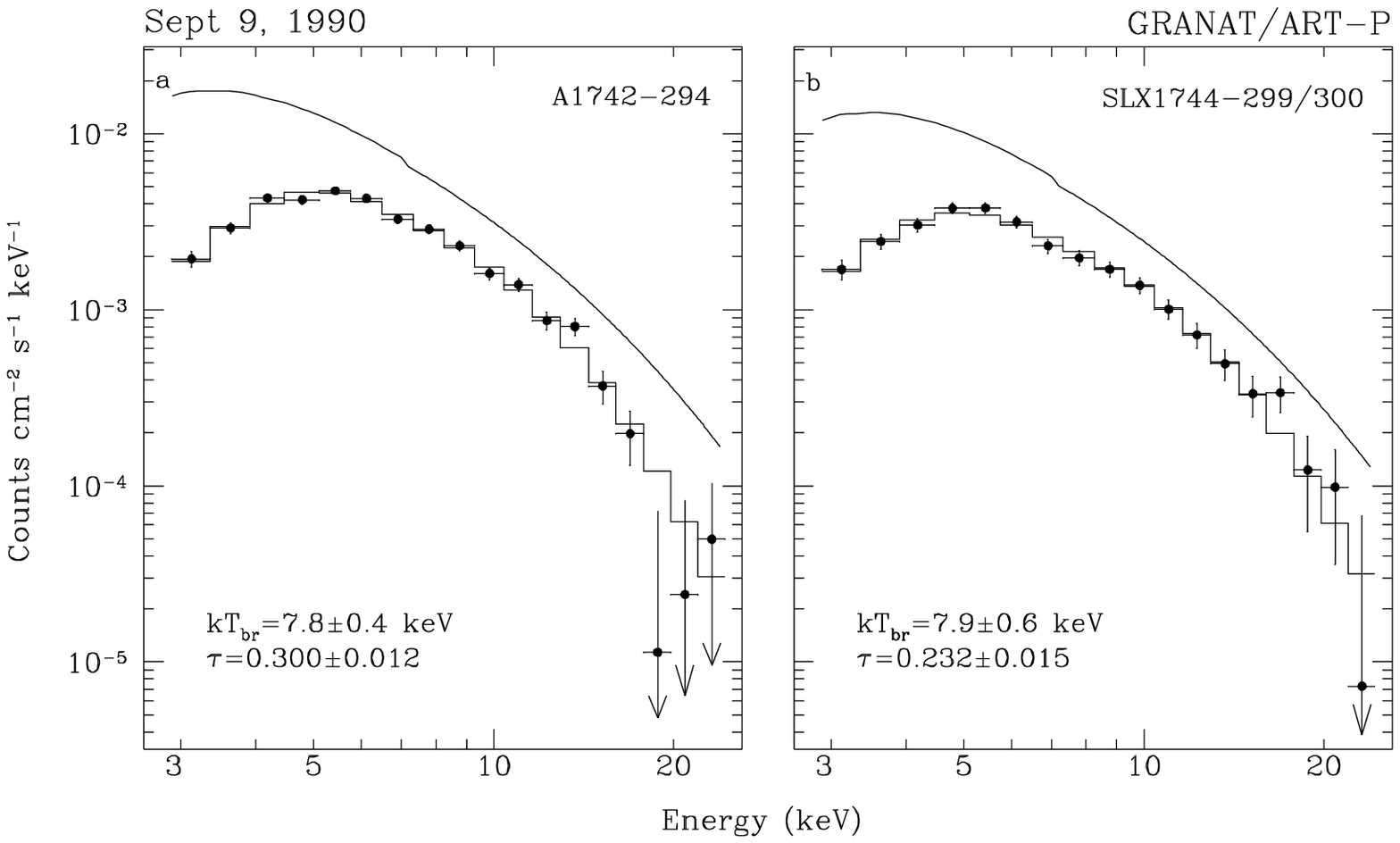}
\end{figure}
 
\vspace{0mm} 
 
\begin{minipage}[]{160mm}{{\vspace{0mm} FIG 1. Spectra of the persistent
emission of A1742-294 (a) and \SLX (b) obtained by GRANAT/ART-P on Sept 8,
1990. The best-fit models are shown by histograms. The solid lines represent
the corresponding photon spectra.}}
\end{minipage}

\pagebreak

\begin{figure}[t]
\epsfxsize=140mm
\epsffile[30 380 502 700]{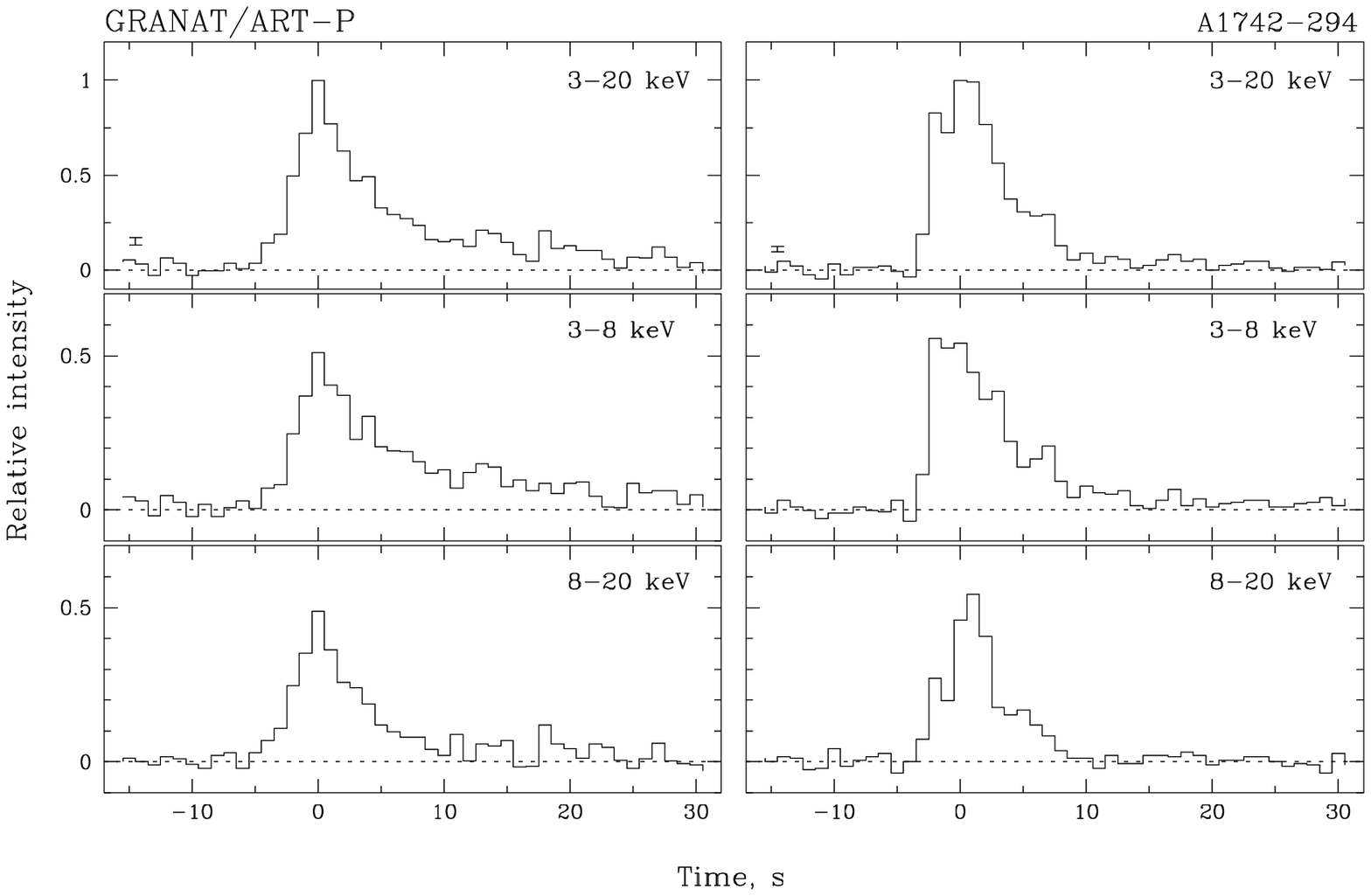}
\end{figure}
 
\hspace{0mm} 
 
\hspace{-5mm}\begin{minipage}[h]{160mm}{{\vspace{0mm} FIG 2. The average
profiles of weak ({\it left panel}) and strong ({\it right panel}) X-ray bursts
detected from A1742-294. }} 
\end{minipage}
 
\vspace{40mm}
 
\begin{figure}[h]
\epsfxsize=70mm
\hspace{27mm}\epsffile[35 365 265 570]{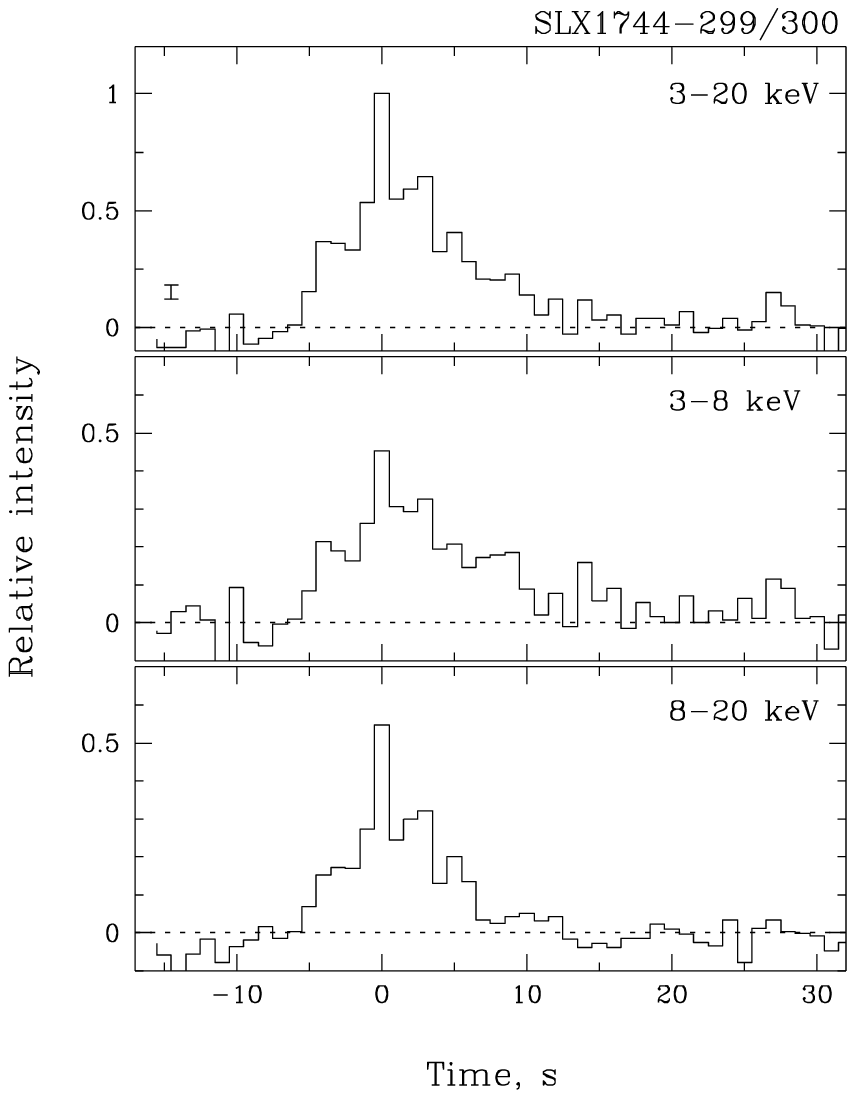}
\end{figure}
 
\hspace{0mm} 
 
\hspace{0mm}\begin{minipage}[h]{160mm}{{\vspace{0mm} FIG.3. The average
SLX1744-299/300 X-ray burst profiles in different energy bands.}} 
\end{minipage}

\pagebreak

\begin{figure}[t]
\epsfxsize=100mm
\vspace{10mm}\hspace{30mm}\epsffile[100 300 490 650]{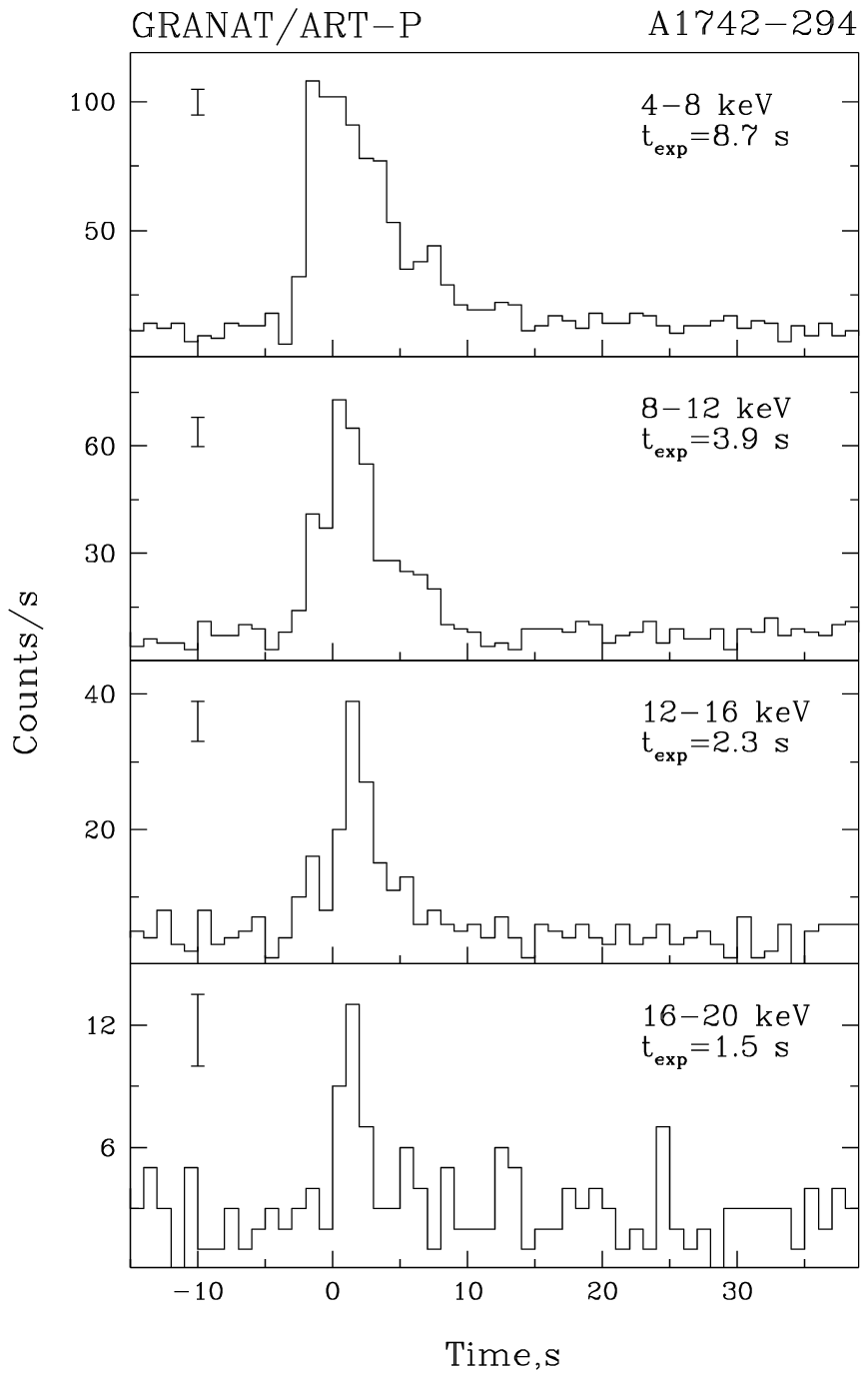}
\end{figure}
 
\begin{minipage}[h]{150mm}{{\vspace{-0mm} FIG.4. Time profiles of the strong
X-ray burst detected by ART-P from A1742-294 on Oct 18, 1990.}} 
\end{minipage}

\begin{figure}[h]
\epsfxsize=85mm
\hspace{30mm}\epsffile[100 415 430 700]{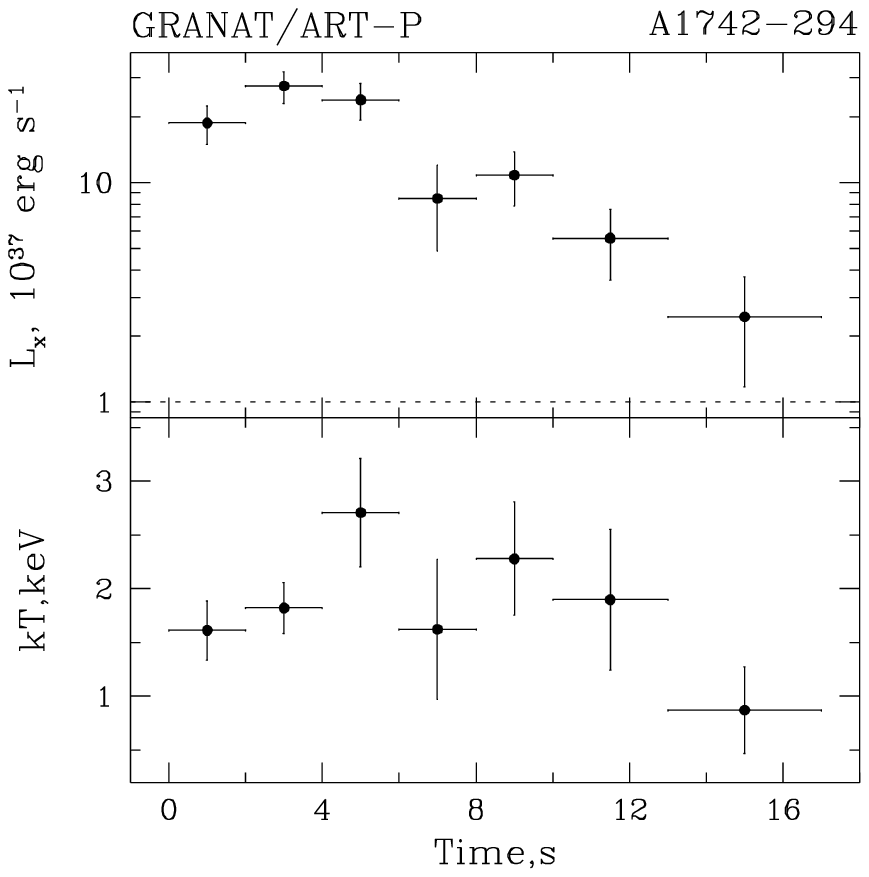}
\end{figure}

\begin{minipage}[h]{160mm}{{\vspace{0mm} FIG.5. Evolution of the
A1742-294 luminosity in 3-20 keV energy band during the Oct 18, 1990 burst
({\it upper panel}). The persistent luminosity level is shown by dotted
line. The evolution of corresponding best-fit black-body temperature is
shown in the {\it bottom panel}. }}
\end{minipage}

\end{document}